\documentclass{aa}

\usepackage[varg]{txfonts}
\usepackage[colorlinks=true, linkcolor=blue, citecolor=blue, urlcolor=gray]{hyperref}
\usepackage{xcolor}
\usepackage{listings}
\usepackage{amsmath}
\usepackage{subcaption}
\usepackage{float}
\usepackage{ulem}

\bibpunct{(}{)}{;}{a}{}{,}

\newcommand{\bs}[1]{\boldsymbol{#1}}

\begin{document} 

\title{Tetrahedral grids in Monte Carlo radiative transfer}

\author{%
Arno Lauwers
\and
Maarten Baes
\and
Peter Camps
\and
Bert Vander Meulen
}

\institute{Sterrenkundig Observatorium, Universiteit Gent, Krijgslaan 281, 9000 Gent, Belgium \newline \email{arno.lauwers@ugent.be}
\label{UGent}
}

 \date{\today}

\abstract
{To understand the structures of complex astrophysical objects, 3D numerical simulations of radiative transfer processes are invaluable. For Monte Carlo radiative transfer, the most common radiative transfer method in 3D, the design of a spatial grid is important and non-trivial. Common choices include hierarchical octree and unstructured Voronoi grids, each of which has advantages and limitations. Tetrahedral grids, commonly used in ray-tracing computer graphics, can be an interesting alternative option.
%The selection of computational grids to discretise equations offers various options, with common choices including the k-d tree or even the unstructured Voronoi grids. Noteworthy developments in literature introduce an efficient ray traversal method tailored specifically for unstructured tetrahedral grids, where the grid cells are tetrahedra. These grids are also commonly applied in computational fluid dynamics and in some astrophysical hydrodynamical simulations.
}
{We aim to investigate the possibilities, advantages, and limitations of tetrahedral grids in the context of Monte Carlo radiative transfer. In particular, we want to compare the performance of tetrahedral grids to other commonly used grid structures.}
{We implemented a tetrahedral grid structure, based on the open-source library TetGen, in the generic Monte Carlo radiative transfer code SKIRT. Tetrahedral grids can be imported from external applications or they can be constructed and adaptively refined within SKIRT. We implemented an efficient grid traversal method based on Pl\"ucker coordinates and Pl\"ucker products.}
{The correct implementation of the tetrahedral grid construction and the grid traversal algorithm in SKIRT were validated using 2D radiative transfer benchmark problems. Using a simple 3D model, we compared the performance of tetrahedral, octree, and Voronoi grids. With a constant cell count, the octree grid outperforms the tetrahedral and Voronoi grids in terms of traversal speed, whereas the tetrahedral grid is poorer than the other grids in terms of grid quality. All told, we find that the performance of tetrahedral grids is relatively poor  compared to octree and Voronoi grids.}
{Although the adaptively constructed tetrahedral grids might not be favourable in most media representative of astrophysical simulation models, they still form an interesting unstructured alternative to Voronoi grids for specific applications. In particular, they might prove useful for radiative transfer post-processing of hydrodynamical simulations run on tetrahedral or unstructured grids.}


\maketitle

%%%%%%%%%%%%%%%%%%%%%%%%%%%%%%%%%%%%%%%%%%%%%%%%%%%%%%%%%%%%%%%%%%%

\section{Introduction}

Nearly all our astronomical knowledge is based on the observation and detailed analysis of radiation we obtain from astronomical sources over the electromagnetic spectrum. Through processes like absorption, scattering, and re-emission, this radiation is affected and altered by the medium it passes through. As a result, detailed modelling of these interactions, that is, radiative transfer simulations, is required. This is particularly important if we want to generate synthetic observations for snapshots of (magneto)hydrodynamical simulations that can directly be compared to observations.

The 3D radiative transfer problem is a notoriously hard nut to crack, and various approaches have been proposed. The Monte Carlo method \citep{2011BASI...39..101W,  2013ARA&A..51...63S, 2019LRCA....5....1N} is the most widely used method for various types of radiative transfer problems. The method is intuitive, conceptually straightforward,  and, contrary to other methods such as finite difference techniques, easily applicable to 3D problems. In the past few years, advanced and powerful Monte Carlo radiative transfer codes have been developed for a variety of radiative transfer applications, such as dust, resonant line, atomic and molecular line, X-ray, UV ionising radiation, and neutrino radiative transfer \citep[e.g.][to name just a few generic examples]{2006A&A...459..797P, 2012ascl.soft02015D, 2016A&A...593A..87R, 2019A&C....27...63H, 2020MNRAS.494.1919L, 2020A&C....3100381C, 2023ascl.soft06034S}.

Three-dimensional Monte Carlo radiative transfer codes generally need a grid onto which the medium is discretised. This grid contains a large number of cells within which all physical properties are considered uniform. Early Monte Carlo radiative transfer simulations often employed simple grids such as spherical shells or regular Cartesian grids. For modern applications, and in particular for the radiative transfer post-processing of hydrodynamical simulations, this approach is not ideal, as these simulations typically cover a large dynamic range in densities. The standard approach to dealing with this complexity is to use hierarchical Cartesian grids, such as octree or binary trees. Many Monte Carlo radiative transfer codes are equipped with such grids \citep{2001A&A...379..336K, 2013A&A...554A..10S, 2014A&A...561A..77S, 2019A&A...622A..79J}.

More recently, unstructured Voronoi grids have gained popularity. Several Monte Carlo radiative transfer codes have now been equipped with Voronoi grids \citep{2013A&A...560A..35C, 2016MNRAS.456..756H, 2017ApJS..233....1K, 2020ApJ...905...27S, 2021MNRAS.506.5129B, 2021A&A...647A..27T}. The use of Voronoi grids in radiative transfer, and in computational methods in general, has several advantages. Voronoi meshes have the ability to deal with arbitrary complex geometries with a large dynamical range, without any preferential directions. Since a Voronoi grid is completely determined by the position of the generating nodes, it is easy to adapt the cell sizes and grid resolution according to user-defined criteria. In general, Voronoi grids allow the dynamical range in densities to be closely reflected  with as few cells as needed.

Voronoi grids are not the only type of unstructured meshes used in computational geometry. Tetrahedral meshes --- tessellations in which all cells are tetrahedra --- are an interesting alternative. Tetrahedral grids are very popular in computer graphics and 3D reconstruction applications \citep[e.g.][]{Georgii2006, MS2006, maria:hal-01486575, Morrical2023}. A tetrahedron is the most elementary polyhedron: each tetrahedron has only four vertices and six edges, and is bounded by four triangular faces. In comparison, an average Voronoi cell in a 3D Voronoi tessellation has approximately 27 vertices and 41 edges, and is bounded by approximately 16 convex polygonal faces \citep{1994A&A...283..361V}. Tetrahedral grids are therefore `simpler' than Voronoi grids and more economical in terms of storage.

In addition to their ability to efficiently represent media with a large dynamical range, the most important characteristic of grids in the context of  Monte Carlo radiative transfer is whether grid traversal can be implemented in a robust and effective manner. Grid traversal essentially comes down to determining, for a path defined by a starting position and direction, the ordered list of cells crossed along this path and the individual path length contributions covered within each individual cell. In many Monte Carlo radiative transfer simulations, grid traversal is responsible for a large fraction of the computational time. 

For Voronoi grids, grid traversal is non-trivial and requires the calculation of the intersection points of the path and each of the faces of the cell \citep{2013A&A...560A..35C, 2016MNRAS.456..756H}. One might expect that grid traversal through a tetrahedral grid would be more efficient than through a Voronoi grid. A tetrahedron has only four faces, so even using a naive algorithm in which we calculate the intersection points of the path and each of the faces, it can probably be quite efficient (remember that Voronoi cells have, on average, almost four times as many faces). Moreover, efficient grid traversal algorithms have been developed for tetrahedral grids in the field of computer graphics \citep{PNT2003, MS2006, LD2008, maria:hal-01486575}.

In this work, we explored the use of tetrahedral grids in Monte Carlo radiative transfer. More specifically, we implemented tetrahedral grids and an advanced grid traversal algorithm in the Monte Carlo radiative transfer code SKIRT \citep{2015A&C.....9...20C, 2020A&C....3100381C}. In Sects.~\ref{sec:GridConstruction} and \ref{sec:GridTraversal} we discuss the construction and traversal of tetrahedral grids in SKIRT, and in Sect~\ref{sec:Tests} we validate our implementation using established benchmark models. We then compare the efficiency of tetrahedral and other grids in Sect.~\ref{sec:Comparison} and discuss these results in a broader context in Sect.~\ref{sec:Discussion}. Finally, we present our conclusions in Sect.~{\ref{sec:Conclusions}}.

\section{Tetrahedral grid construction in SKIRT}
\label{sec:GridConstruction}    
    
\subsection{The SKIRT radiative transfer code}

SKIRT \citep{2015A&C.....9...20C, 2020A&C....3100381C} is a generic 3D Monte Carlo radiative transfer code. Originally conceived as a dust radiative transfer code \citep{2003MNRAS.343.1081B, 2011ApJS..196...22B}, it has transformed into a more generic tool that can be applied to various radiative transfer applications. Apart from dust radiative transfer, it can now perform Ly$\alpha$ resonant line radiative transfer \citep{2021ApJ...916...39C}, non-local thermodynamic equilibrium line radiative transfer for selected ions, atoms, and molecules \citep{2023A&A...678A.175M}, X-ray radiative transfer \citep{2023A&A...674A.123V}, and H{\sc{i}} radiative transfer \citep{2023MNRAS.521.5645G}. The efficiency of the code is maximised through many optimisation mechanisms and a hybrid parallelisation strategy \citep{2008MNRAS.391..617B, 2013ARA&A..51...63S, 2016A&A...590A..55B, 2022A&A...666A.101B, 2017A&C....20...16V}.

The SKIRT code is equipped with a diverse suite of options to discretise the medium. Apart from the standard spherical, cylindrical and Cartesian grids, the user can select octree grids \citep{2013A&A...554A..10S}, $k$-d tree grids \citep{2014A&A...561A..77S}, and Voronoi grids \citep{2013A&A...560A..35C}. The availability of these different options makes SKIRT a suitable environment to investigate the accuracy and efficiency of tetrahedral grids. 
    
\subsection{Delaunay triangulation}
\label{seq:DT}
        
\begin{figure}
\centering
\includegraphics[width=\linewidth]{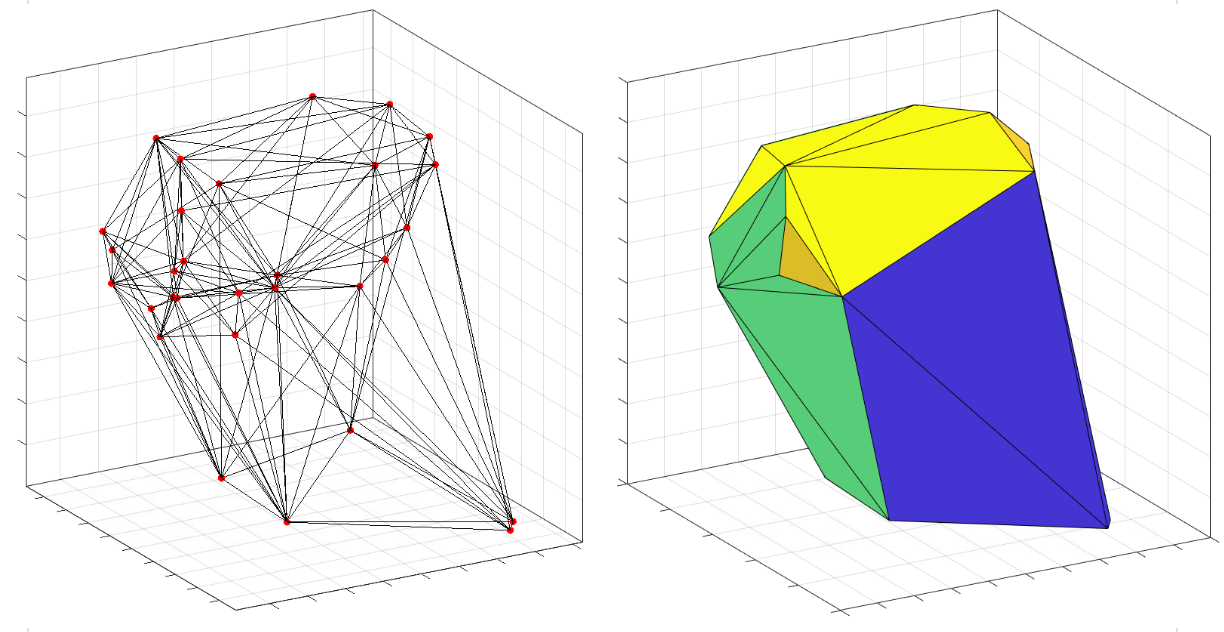}
\caption{ DT of a set of 30 random points drawn from a uniform distribution.}
\label{fig:convex_hull}
\end{figure}

A triangulation of a set of points is a subdivision of the convex hull of those points into non-overlapping simplices. In 3D this is sometimes called a tetrahedralisation as the simplices are tetrahedra. The Delaunay triangulation (DT) of a set of 3D points $\{\bs p_i\}$ is the set of non-overlapping tetrahedra whose vertices are points, $\bs p_i$, that do not lie inside the circumsphere of any tetrahedron in the triangulation. This last condition is also called the Delaunay condition. This ensures a unique triangulation if and only if no more than four points lie on the same sphere. However, even if five or more points lie on the same sphere, we can usually still find a non-unique triangulation. An example of a DT can be seen in Fig.~{\ref{fig:convex_hull}}.
        
Generating the DT from a set of given input of points is a well-known problem in computational geometry \citep{BOROUCHAKI1995153, CIGNONI1998333}. The 2D DT has the useful property that it maximises the smallest angle in each triangle, resulting in regular triangles with nearly square bounding boxes. On the other hand, the 3D DT does not have this property; cells tend to be elongated and very irregular \citep{musin1997properties, chew1997guaranteed, maur2002delaunay}. For many applications an isotropic grid is preferred, making regular tetrahedra the more desirable choice. In order to improve the quality of a tetrahedral grid one can move the vertices, change the Delaunay condition, or add so-called Steiner points in appropriate locations. Several techniques have already been developed for this purpose, such as Delaunay refinement \citep{Shewchuk1970}.

\subsection{TetGen}
        
Given the intricacy of the problem, we chose to use the third-party library TetGen \citep{Si2015} for the construction of tetrahedral grids in SKIRT. TetGen is a {\tt{C++}} quality tetrahedral mesh generator that is widely used in computational fluid dynamics simulations \citep[e.g.][]{shirokov2021mesh, mayer2009interface, bonfiglioli2013unstructured}.

TetGen uses a constrained Delaunay refinement algorithm \citep{tetgen_algo} and offers various features for creating a robust computational grid to perform radiative transfer on. First of all is the ability to generate constrained triangulations, forcing grid cells to adhere to a specified input surface. In our case this would typically be the simulation's bounding box. This means that the simulation's domain is fully covered with grid cells, rather than only partially by the convex hull of the triangulation. Additionally, and most importantly, TetGen supports a mesh sizing function, allowing users to define a function in the grid space that determines the cell refinement. An example of this could be a scalar density field to determine the desired resolution. In practice, TetGen achieves this by letting the user set a function {\tt{tetunsuitable}}, which should return {\tt{true}} if the tetrahedron must be optimised. This is usually done by applying various local mesh operations, such as adding Steiner points. The {\tt{tetunsuitable}} function can represent any condition chosen by the user, such as a maximum ratio of edge lengths, volume, or dihedral angle. If desired, it could also be made anisotropic. It should also be noted that TetGen performs additional relaxing and smoothing operations after applying the mesh sizing function to the entire domain. As a result, the user-defined conditions are not strictly met in all cells. While these operations can be disabled, doing so typically results in a lower-quality grid.

\subsection{TetGen in SKIRT}
\label{sec:TetGen in SKIRT}

\begin{figure}
\centering
\includegraphics[width=\linewidth]{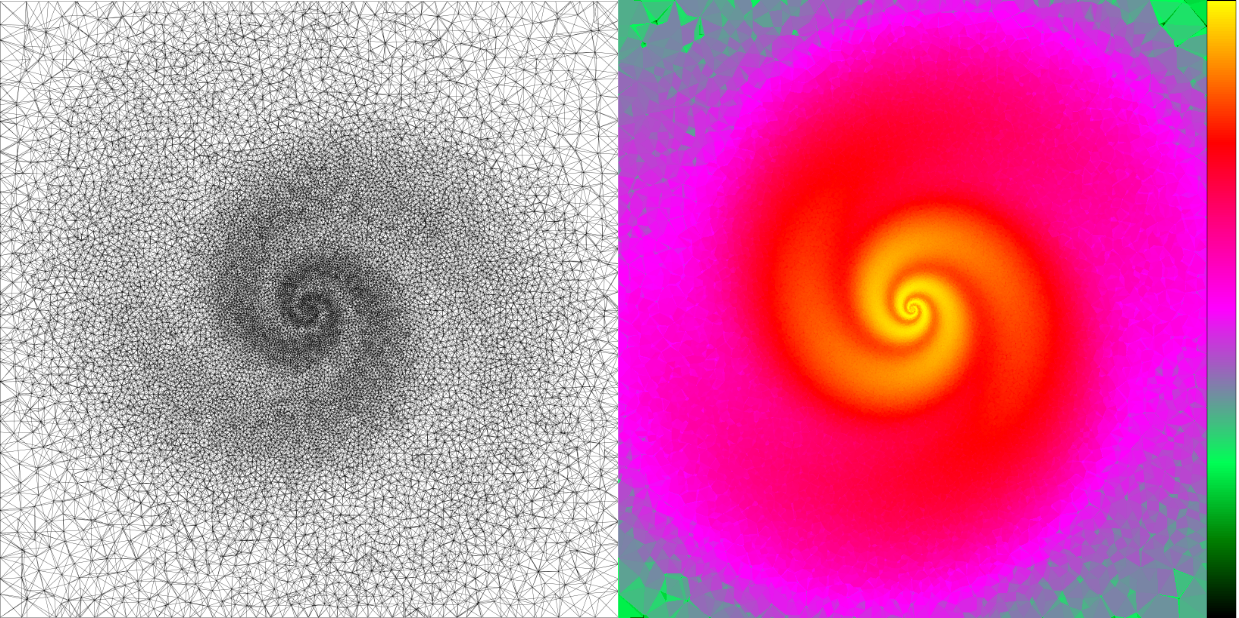}
\caption{TetGen's mesh sizing function applied in SKIRT. Right: Cut through the density field for a dusty spiral model. Left: Cut through the triangulation built using TetGen by constraining the minimum mass fraction per tetrahedron to $10^{-5}$.}
\label{fig:mesh_sizing}
\end{figure}

For our specific purpose of discretising a given density field in the SKIRT radiative transfer code, we needed a recipe on which the refinement algorithm is based. Mimicking the strategy applied for hierarchical grids in SKIRT \citep{2013A&A...554A..10S, 2014A&A...561A..77S}, we chose limiting the total mass fraction inside a cell as our default option. This naturally results in a higher resolution in regions with a larger density. An example of this adaptive grid construction is displayed in Fig.~{\ref{fig:mesh_sizing}}. We employed an additional subdivision criterion, which is also used in the hierarchical tree grids in SKIRT as described by \citet{2014A&A...561A..77S}. We subdivided the cell if the quantity
\begin{equation}
    q = 
    \begin{cases}
        \;\dfrac{\rho_{\text{max}} - \rho_{\text{min}}}{\rho_{\text{max}}} & \text{if } \rho_{\text{max}} > 0, \\
        \;0 & \text{if } \rho_{\text{max}} = 0.
    \end{cases}
\end{equation}
exceeds some user-defined threshold $q_{\text{max}}\le 1$. In this expression, $\rho_{\text{min}}$ and $\rho_{\text{max}}$ represent the minimum and maximum values of the density in the cell, which are estimated by generating a large number of random positions in the cell (see Sect.~{\ref{sec:randompoints}}). This additional criterion avoids the use of cells with large density gradients.

In addition to adaptively constructing the grid, tetrahedral grids can be directly imported from hydrodynamical simulations or generated using other tools. Furthermore, when supplied with particle data, such as from a smoothed-particle hydrodynamic simulation, TetGen can generate the corresponding DT of the point set. The uniform density in each cell can then be derived by interpolating the input densities at the vertices.

\subsection{Generating random points}
\label{sec:randompoints}

After the construction of a grid, the properties of each cell, such as the mass density, must be determined. In the SKIRT radiative transfer code, this is achieved by sampling a user-defined number of random points in each cell and averaging the underlying density at these locations. This requires an algorithm to efficiently sample random points in an arbitrary tetrahedron. A simple option could consist of generating random points inside the bounding box of the tetrahedron and subsequently checking whether they fall inside the tetrahedron. However, more efficient techniques for generating uniformly distributed random points within a tetrahedron exist, and we used the algorithm described by \citet{vc-lab-tetrahedron}.

\section{Grid traversal}
\label{sec:GridTraversal}

% talk about tetrahedron orientation? there are only 2 and TetGen is consistent

\subsection{Pl\"ucker coordinates}
       
 Pl\"ucker coordinates offer a method for representing directed lines in 3D space using 6D vectors \citep{shoemake1998plucker, PNT2003}. Given a directed line $R$ defined by a point $\bs r$ and a direction $\bs k$, its Pl\"ucker coordinates can be expressed as
\begin{align}\label{eq:plucker}
\bs \pi_R=\{\bs k : \bs k \times \bs r\} = \{\bs U_R : \bs V_R\}.
\end{align}
        
We can now define the permuted inner product or Pl\"ucker product of two directed lines $R$ and $S$ as
\begin{align}
\bs \pi_R \odot \bs \pi_S = \bs U_R \cdot \bs V_S + \bs U_S \cdot \bs V_R.
\end{align}
The sign of this product determines the relative orientation between two directed lines. We have
\begin{equation}
\begin{cases}
\;\bs \pi_R \odot \bs \pi_S > 0 &\iff \;\textrm{$S$ goes anti-clockwise around $R$,}\\
\;\bs \pi_R \odot \bs \pi_S < 0 &\iff \;\textrm{$S$ goes clockwise around $R$,}\\
\;\bs \pi_R \odot \bs \pi_S = 0 &\iff \;\textrm{$S$ intersects or is parallel to $R$.}
\end{cases}
\end{equation}
This allows us to efficiently find the exit face when tracing a photon path through a tetrahedron, as will be discussed in the next section.

\subsection{Traversal algorithm}

We use the term `ray' to indicate a given photon path represented by its Pl\"ucker coordinates. Considering the edges of a tetrahedron as directed lines from one vertex to another, a ray exits a face if the clockwise-ordered\footnote{Viewed from the fourth vertex or also from inside.} edges yield a negative Pl\"ucker product with the ray. In other words, if the Pl\"ucker products computed around clockwise-ordered edges move clockwise around the ray, we have an exiting ray. Conversely, if they are all anti-clockwise, indicated by positive Pl\"ucker products, we have an entering ray.

A naive approach would be to calculate these three Pl\"ucker products for each of the four faces. However, despite the edges having a different orientation when associated with each face, there is no need to recalculate the products, as flipping the orientation of the edge simply flips the sign of the resulting Pl\"ucker product. We can thus reduce the number of required Pl\"ucker products to six, one for each edge. This approach was applied by \citet{PNT2003}. However, we usually know the entry face of the ray and thus only need to check the other three faces. This, with some additional optimisations, was used by \citet{MS2006} to reduce the number of Pl\"ucker products to $2.67$ on average. Other approaches exist, such as using three to six scalar triple products, as described in \citet{LD2008}.
        
\begin{figure}
\centering
\includegraphics[width=\linewidth]{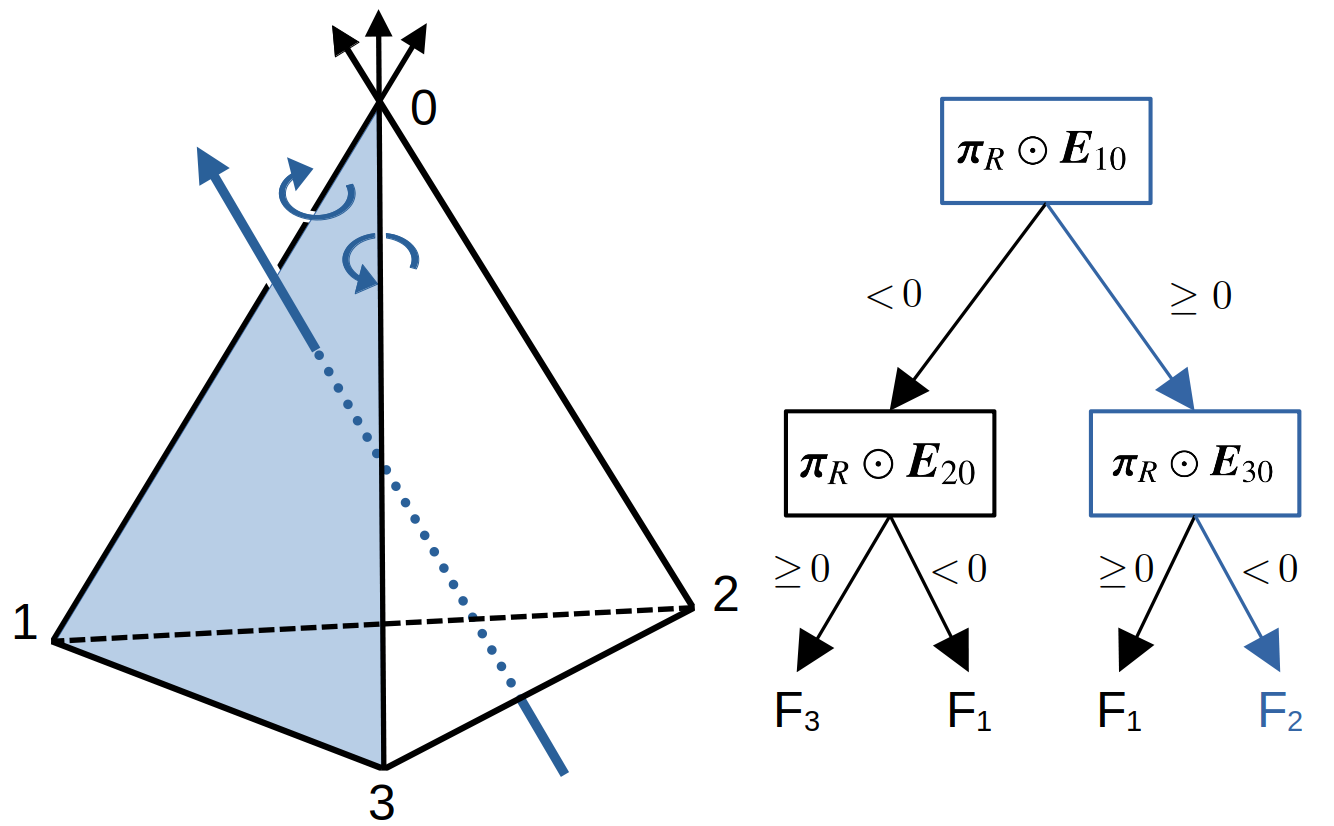}
\caption{Algorithm, as described by \citet{maria:hal-01486575}, that finds the exit face (coloured) using only two Pl\"ucker products (left) and the corresponding decision tree (right). The leaf nodes in the tree are the possible exit faces, numbered according to the opposite vertices.}
\label{fig:treealgo}
\end{figure}

On the other hand, \citet{maria:hal-01486575} described a very efficient method that only requires two Pl\"ucker products. This is illustrated in Fig.~\ref{fig:treealgo}. We label the faces using the opposite vertices and use face $0$ as the entry face. We started by computing the Pl\"ucker product of the edge $1\rightarrow 0$ with the ray. If this results in a (anti-)clockwise orientation, we next used the edge that is (anti-)clockwise viewed from vertex $0$. We computed the Pl\"ucker product of this new edge and the ray. If these two calculated Pl\"ucker products have an opposite sign, then the exit face is the face bounded by these two edges. This can be understood if we remember that the edges of the entry face have a known orientation with the ray\footnote{Anti-clockwise orientation with clockwise ordered edges (opposite from the leaving ray).}, as it is the entering face. Using this additional information for the last edge, means we have found three edges with a clockwise orientation in a clockwise order.
If instead they have an equal sign, then the remaining edge, which we do not check, will have an opposite-signed Pl\"ucker product. This will make face $1$ the exit face, which can be understood with an argument analogous to the above.
%\end{itemize}

Since all the possible edges we have to check intersect at the same point, vertex $0$, we can shift our coordinate frame such that the origin coincides with this vertex. This will speed up the computation by ensuring the Pl\"ucker coordinates, \eqref{eq:plucker}, of the edges have a vanishing moment $\bs V_E$. We will, however, need to shift the ray to this new coordinate system by updating its moment to
\begin{align}
\bs V_R = \bs k \times \left(\bs r - \bs v_0\right),
\end{align}
where $\bs v_0$ are the coordinates of vertex $0$. Once the exit face is found we can find the distance to this face using
\begin{align}\label{eq:plane_dist}
s=\dfrac{\bs n \cdot \left(\bs r - \bs v_0\right)}{\bs n \cdot \bs k}
\end{align}
where $\bs n$ is normal to the intersected face. We can then move the ray forwards by a distance $s$.

However, when starting the algorithm we did not have the entering face information. Thus, for the first step, we needed to take a slightly different approach, which was also suggested by \citet{maria:hal-01486575}. Once we determined the tetrahedron containing the ray position, we calculated the Pl\"ucker product with an arbitrary edge. This orientation indicates which of the two adjacent faces cannot be the exit face. Subsequently, we can apply the previous algorithm, using this face as the `fake' entry face. The algorithm will still work even if this face is not the true entering face.

We should also mention that there are a few (rarely occurring) edge cases, such as the ray going through a vertex or edge, that need to be handled with care. We used a simple plane intersection test whenever the leaving face might be ambiguous to resolve these edge cases.

\subsection{Locating the ray origin}

A Monte Carlo radiative transfer simulation differs from a traditional ray-tracing simulation in that each ray's origin is generated randomly, so that the cell containing this position needs to be determined every time a ray is fired. To accomplish this, we adopted a strategy inspired by \citet{2013A&A...560A..35C} and constructed a $k$-d search tree using the centroids of the tetrahedra. When we wanted to find the cell containing a given point, we located the closest centroid using the $k$-d tree and used our traversal algorithm to traverse from the found centroid towards the given input point. It is important to note that the nearest centroid often does not belong to the tetrahedron containing the input point. If, during traversal, the distance travelled exceeds the distance between the found centroid and the input point, we have found the desired tetrahedron. In the very rare event the algorithm fails, we can resort to a loop to check all tetrahedra. 

It is also a viable option to use a breadth-first search over the neighbours around the found cell. We conducted several tests and found that using the traversal algorithm is very reliable and the most efficient.

\section{Validation}
\label{sec:Tests}

To verify our implementation, we performed two dust radiative transfer benchmarks, comparing our results with those calculated using a regular cylindrical SKIRT grid as well as with results from other radiative transfer codes.

\begin{figure*}
    \centering
    \includegraphics[width=0.47\textwidth]{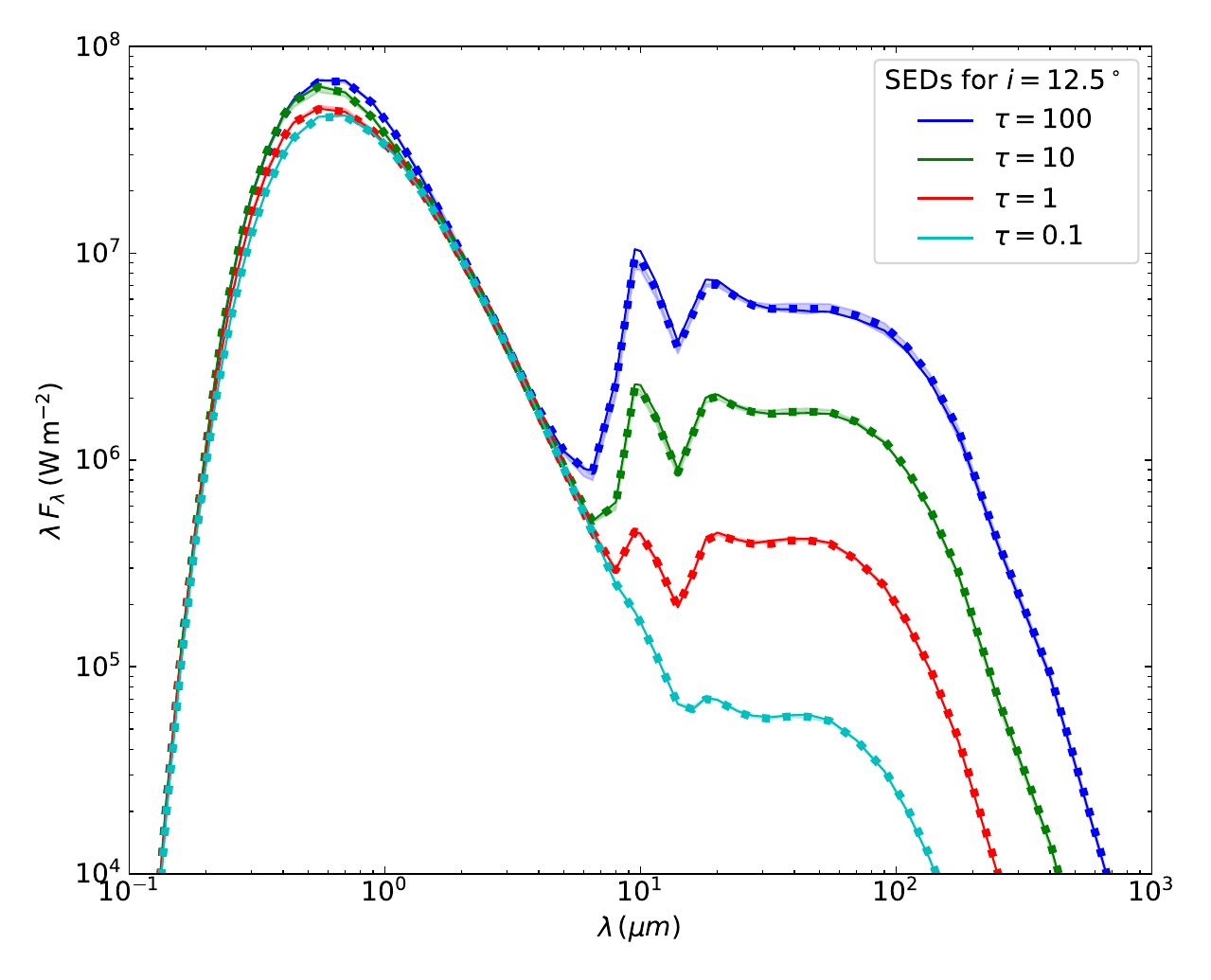}
    \hfill
    \includegraphics[width=0.47\textwidth]{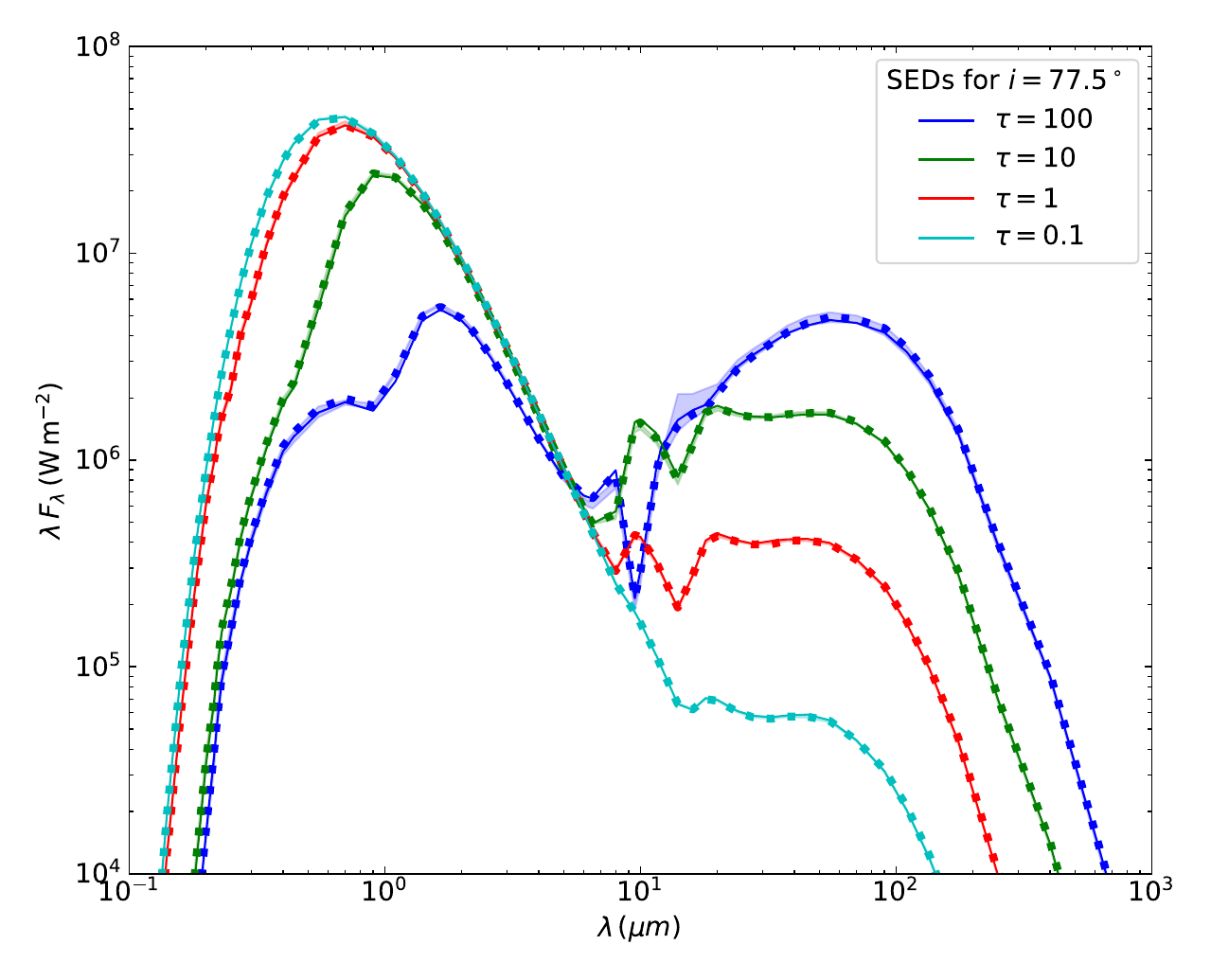}
    \caption{Results from the benchmark described by \citet{2004A&A...417..793P}. The simulation was performed on both a regular (cylindrical) and a tetrahedral grid in SKIRT and using six other radiative transfer codes. The figure shows the SED for two inclinations, $i=12.5^\circ$ and $i=77.5^\circ$, relative to the disc, and for optical depths of $0.1$, $1$, $10$, and $100$. The solid and dotted lines represent the results using our grid and the regular grid in SKIRT, respectively. The coloured areas represent the minimum and maximum from the other radiative transfer codes' benchmark results.}
    \label{fig:Pascucci}
\end{figure*}

\textbf{Pascucci et al.\ (2004):}
The \citet{2004A&A...417..793P} benchmark features a central T~Tauri star embedded in a circumstellar disc with an inner cavity free of dust. The density structure of the disc represents that of a massless Keplerian disc. The challenge of the radiative transfer problem is that the density distribution provides a significant density gradient in the inner part of the disc. The benchmark performs a panchromatic simulation and focuses on the spectral energy distributions (SEDs); it is thus less sensitive to the individual paths taken and will validate other aspects, such as calculated path lengths and absorption. In the original work, five different radiative transfer codes participated to the benchmark, and all of them reproduce the SEDs very well, with relative differences generally below 10\%, except for the most challenging cases (high optical depth and edge-on orientation). We ran the benchmark simulations with SKIRT, using both tetrahedral and regular cylindrical grids, with $30$ and half a million cells, respectively. As shown in Fig.~{\ref{fig:Pascucci}}, our results reproduce the benchmark results very well, with an absolute relative error of $1.5\%$ and $2\%$ for $i=12.5^\circ$ and $i=77.5^\circ$, respectively.

\begin{figure*}
    \centering
    \includegraphics[width=\textwidth]{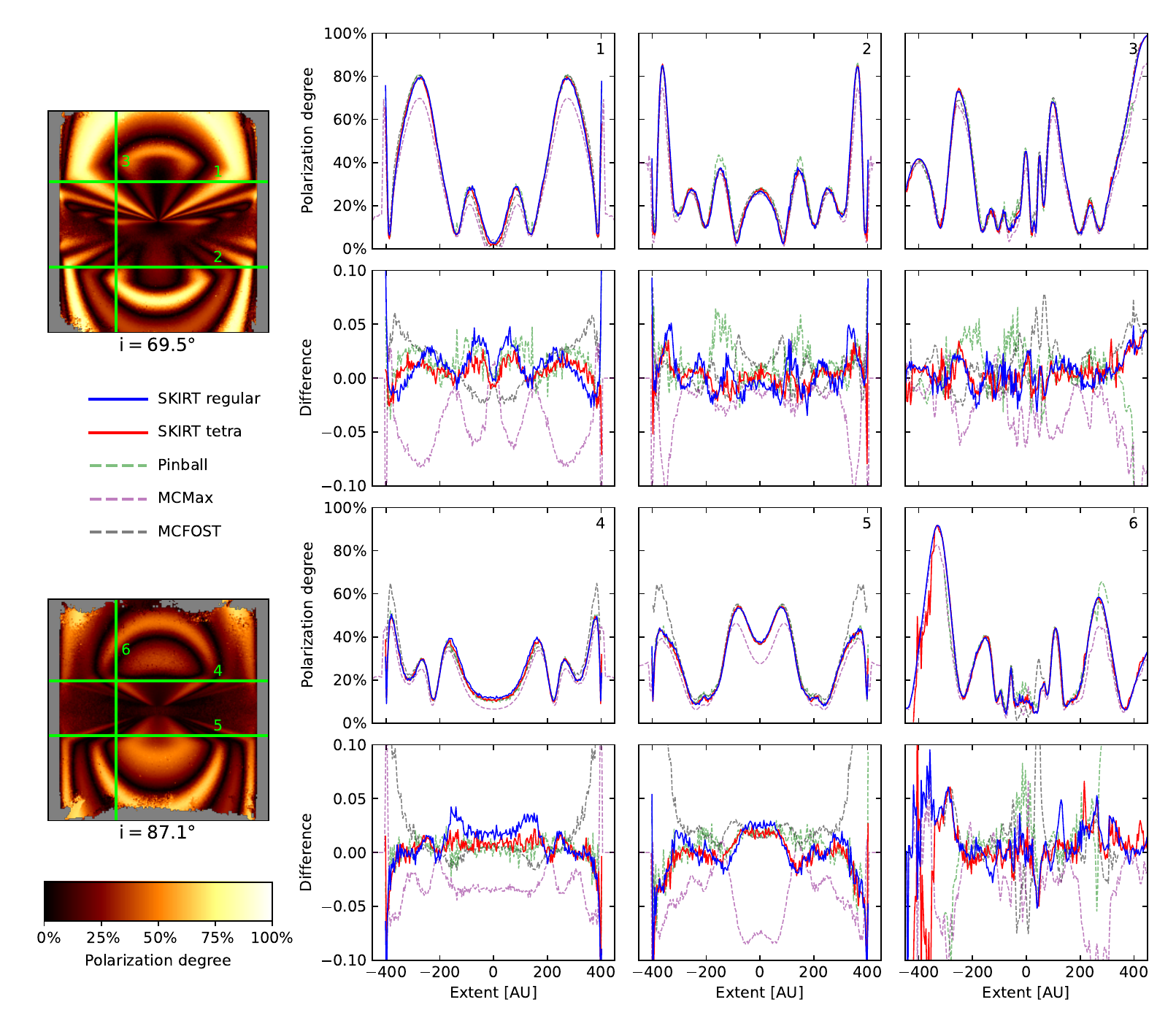}
    \caption{Results from the benchmark described by \citet{2009A&A...498..967P}. The simulation was performed on both a regular cylindrical and a tetrahedral grid in SKIRT and using three other radiative transfer codes. The polarisation degree and the difference to the means of all benchmark results are shown for the two inclinations $i=69.5^\circ$ and $i=87.1^\circ$ relative to the disc.}
    \label{fig:Pinte}
\end{figure*}

\textbf{Pinte et al.\ (2009):}
The \citet{2009A&A...498..967P} benchmark builds on the \citet{2004A&A...417..793P} benchmark but extends it to more challenging values. The optical depth extends to larger values (up to $10^6$ in the midplane), scattering is anisotropic instead of isotropic, and scattering polarisation is included. With images using SKIRT polarisation \citep{2017A&A...601A..92P}, this benchmark provides a very strong test for the accuracy of the individual photon paths. The results for this benchmark are depicted in Fig.~\ref{fig:Pinte}. The SKIRT results corresponding to the cylindrical and tetrahedral grids are in close agreement, with an average error of only $0.3\%$ and $0.5\%$ for $i=69.5^\circ$ and $i=87.1^\circ$, respectively. The tetrahedral and cylindrical grid respectively used $41$ and $6$ million cells.
The artefacts observed in the outer regions of the polarisation maps stem from using a photon count that was not scaled with the amount of cells required by our 3D grid. When attempting to accurately represent the 2D medium with a 3D grid, a significantly greater number of cells was necessary compared to the larger cylindrical cells. However, we did not proportionately increase the photon count, resulting in fewer photons per cell within the 3D grid. Consequently, in the outer regions where the flux is typically 20 orders of magnitude lower than in the central region, some cells end up with zero flux, leading to an undefined polarisation degree in those areas.

\section{Comparison to other grids}
\label{sec:Comparison}    

We next compared the performance of the tetrahedral grid to that of the unstructured Voronoi grid and the hierarchical octree grid already present in SKIRT. To this end, we constructed a simple but fully 3D model, and we ran simulations with various grid types and cell counts. We measured the simulation run time and evaluated the grid quality using a metric based on synthetic observations of this model.

\begin{figure}
    \centering
    \includegraphics[width=\linewidth]{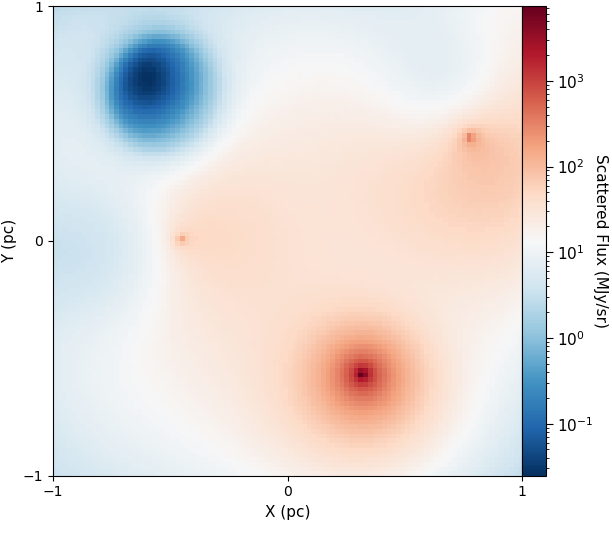}
    \caption{Image of the scattered flux from our test model with ten Plummer spheres of dust and three point sources. The dynamic range of the flux is close to six orders of magnitude.}
    \label{fig:plummer}
\end{figure}

The simulation model used for our tests is defined in a cubical spatial domain. The cube contains a dust medium defined by ten superposed, arbitrarily placed Plummer spheres with varying radii and dust masses. The spheres are cut off at the boundaries of the cube. There are also three random point sources placed at arbitrary locations within the cube and thus embedded in the medium. We performed monochromatic simulations that generate parallel-projected 100 by 100 pixel images of this system for three different sight lines. We exclusively used the scattered flux to eliminate the error from the bright point sources in the images. One of these images is shown in Fig.~\ref{fig:plummer}. The dust density in the cube varies by four orders of magnitude, and the optical depth from the centre of the cube to its boundary ranges from 2.5 to 450. While this geometry is quite artificial, for our purposes it should be fairly representative of many astrophysical simulation models.

We ran simulations of the test model with varying cell counts. Constructing a Voronoi grid with a given cell count is straightforward: we can simply specify the number of Voronoi sites to be sampled from the dust distribution. However, tetrahedral and octree grids are created adaptively. For these grid types, we varied the maximum dust mass fraction allowed in a grid cell to control the cell count. A stricter mass fraction limit will cause extra cells, especially in regions with higher density.

\begin{figure*}
    \centering
    \includegraphics[width=0.47\textwidth]{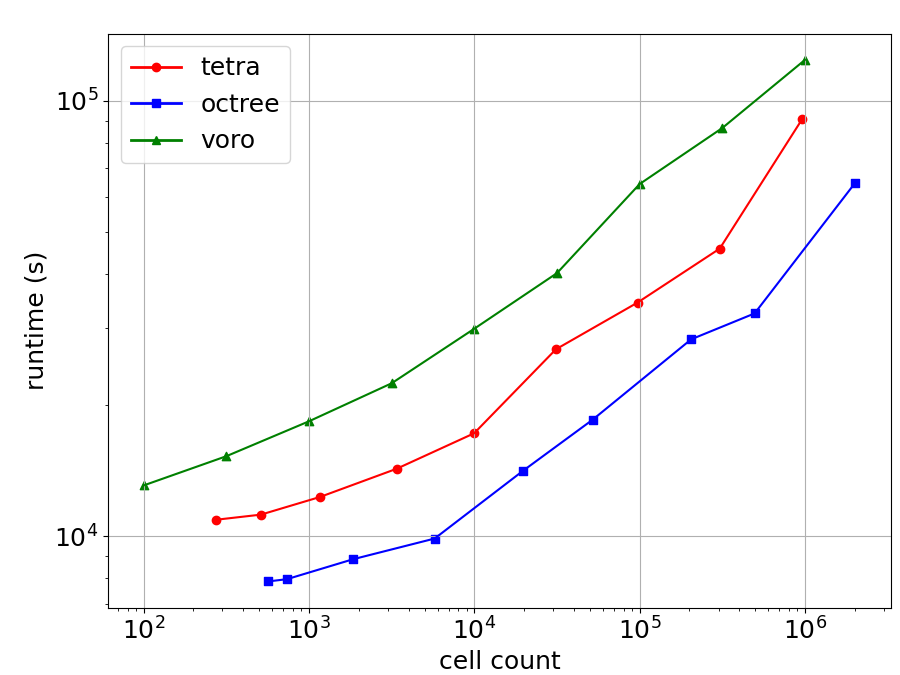}
    \hfill
    \includegraphics[width=0.47\textwidth]{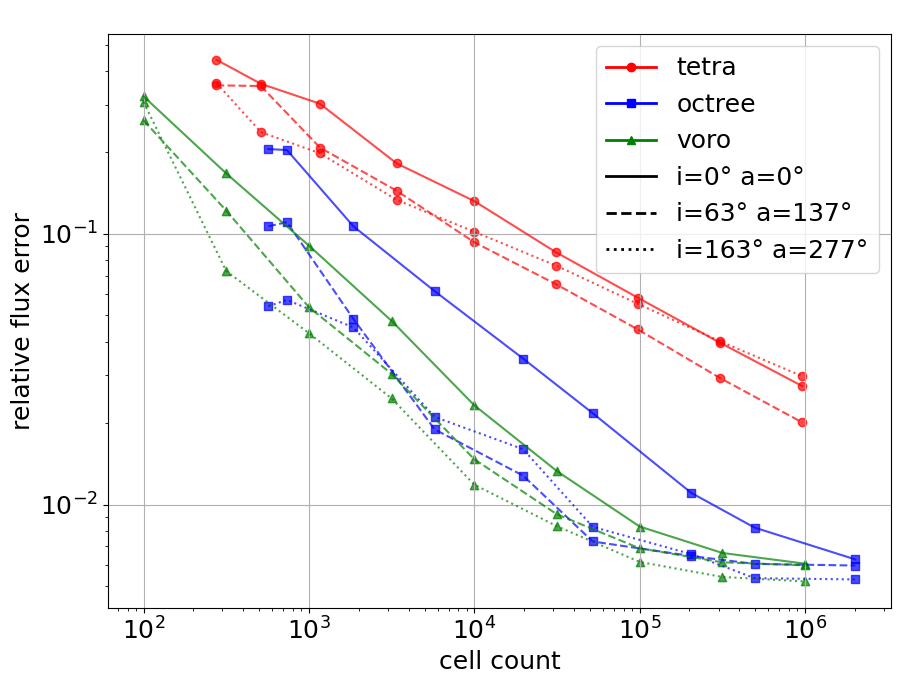}
    \caption{Comparison of the tetrahedral grid with the octree Voronoi and grids. Left: Simulation runtime of our test model as a function of the number of grid cells. Right: Mean of the relative flux residuals, compared pixel by pixel to the corresponding reference image for three sight lines, as a function of the number of grid cells.}
    \label{fig:cell_comparison}
\end{figure*}

The leftmost panel of Fig.~\ref{fig:cell_comparison} shows the runtime of our test model simulations as a function of the number of grid cells. As expected, the slopes of the curves are similar for the three grid types. At constant cell count, the octree grid significantly outperforms the other grids. This is unsurprising given the simple Cartesian cell geometry and in view of the extensively optimised implementation as documented by \citet{2013A&A...554A..10S, 2014A&A...561A..77S}. On the other hand, at constant cell count, the Voronoi grid is substantially slower than the tetrahedral grid. Again, this is not surprising because, compared to tetrahedrons, Voronoi cells have a more complex geometry and on average have more neighbouring cells.

Comparing run time at constant cell count tells only half of the story. Some grid types may require fewer cells, and thus a correspondingly shorter runtime, to achieve the same results. We thus needed a metric to evaluate the `quality' of the spatial grid employed in the simulation. This is ultimately determined by the accuracy of the generated synthetic observables. We therefore defined a quality measure based on the images generated by our test setup. We first produced a reference image for each sight line by running the simulation using a very high-resolution octree grid and a large number of photon packets, ensuring that the fluxes in each pixel are numerically converged. To derive a quality metric for a given test run, we calculated the absolute value of the relative error on the flux in each pixel, as compared to the corresponding reference pixel. The final quality metric is then defined as the mean value of these relative errors for all pixels in the image.

The rightmost panel of Fig.~\ref{fig:cell_comparison} shows this quality metric for the three sight lines of our test model simulations, as a function of the number of grid cells. At constant cell count, the tetrahedral grid produces a significantly higher relative error than the other grid types. Or equivalently, for a given level of accuracy, the tetrahedral grid requires a substantially larger number of cells. Notably, the sight line along the Cartesian axes ($i{=}0^\circ$~$a{=}0^\circ$) shows an overall worse quality, especially for the octree grid. This can be attributed to discretisation effects caused by having both the sight line and all cells in the grid lined up with the coordinate axes. This occurs for both the octree grid and high-resolution octree reference, which explains why this can also be seen in the unstructured grids.

\begin{figure}
    \centering
    \includegraphics[width=\linewidth]{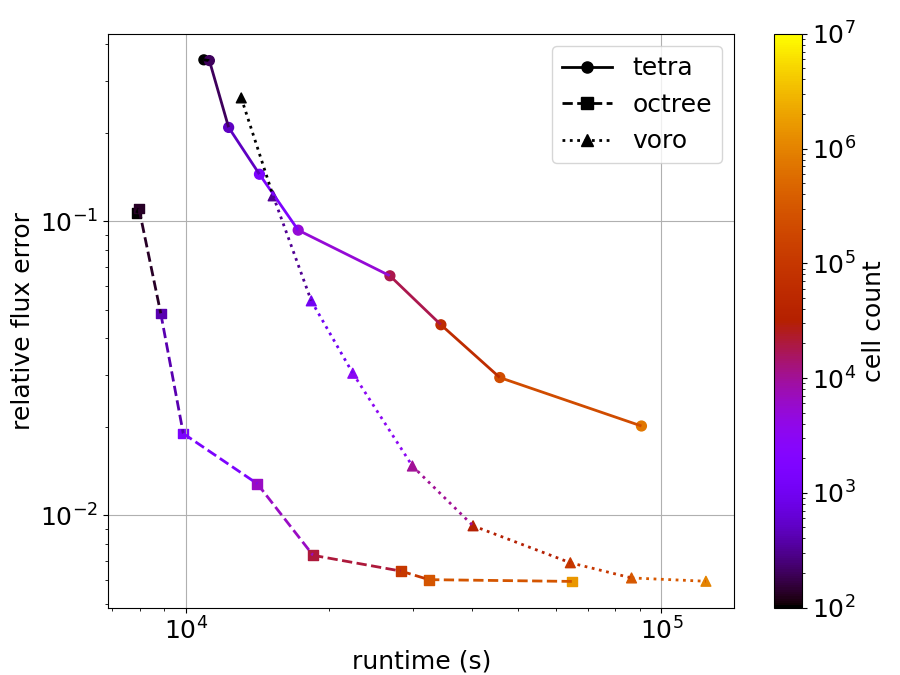}
    \caption{Comparison of the tetrahedral grid with the Voronoi and octree grids, now showing the relative error (the mean of the pixel-by-pixel flux residuals) as a function of simulation runtime. The colour indicates the number of grid cells used.}
    \label{fig:comparison}
\end{figure}

The grid performance is compared more directly in   Fig.~\ref{fig:comparison}, where we plot the relative error as a function of simulation runtime. To avoid an unfair comparison, we excluded the sight line at zero inclination and the azimuth from this summary plot, instead averaging the relative errors for the two other sight lines. For a given runtime, the tetrahedral grid consistently produces a larger relative error than the other grid types, at least in the relevant range where the relative error is below 10 per cent. Compared to the octree grid, the tetrahedral grid requires more cells for the same quality and it runs more slowly for an equal cell count (see Fig.~\ref{fig:cell_comparison}), which causes the substantial overall performance gap shown in Fig.~\ref{fig:comparison}. Compared to the Voronoi grid, the tetrahedral grid runs faster for equal cell count, but this advantage is insufficient to compensate for the extra number of cells required for the same quality.

The conclusions drawn using our smooth test model cannot necessarily be generalised to simulation models with other properties. For example, we expect that unstructured grids will perform much better for models with steep density gradients, resembling hard boundaries such as those often encountered in computational fluid dynamics, where tetrahedral grids are prevalent. This is especially true if the gradients are not lined up with the coordinate axes, putting the Cartesian octree grid at a disadvantage. Furthermore, constructing the tetrahedral grid using other refinement options could also improve its performance.

\section{Discussion}
\label{sec:Discussion}    

\subsection{Hierarchical versus unstructured grids}

Tetrahedral grids form a potentially interesting spatial grid structure for Monte Carlo radiative transfer, with advantages and disadvantages with respect to other grid types. One of the interesting characteristics of tetrahedral grids is that they are unstructured. In this sense, they are similar to Voronoi grids, which have been implemented in several Monte Carlo radiative transfer codes \citep{2013A&A...560A..35C, 2016MNRAS.456..756H, 2017ApJS..233....1K, 2020ApJ...905...27S, 2021MNRAS.506.5129B, 2021A&A...647A..27T}. One of the advantages of unstructured grids is that they are more flexible and adaptable, and the individual grids cells are not aligned along any preferential directions. In other words, unstructured grids provide a tessellation of the medium without the rigidity of a Cartesian grid. 

Another advantage is that they can be made to conform to any desired geometry in such a way that a minimum number of grid cells is required for a given spatially varying resolution. In the context of radiative transfer, this is important for two different reasons. First, many radiative transfer problems require the calculation of the medium's secondary emission. This requires the measurement of the radiation field in every cell with sufficient statistics, which, in turn, requires sufficient photon packets passing through each cell. Secondly, the grid traversal is not the only and not necessarily the most time-consuming aspect of Monte Carlo radiation transfer. Sometimes the calculation of the emissivity of the grid cells can be a substantial computational effort too. A typical example is thermal emission by dust, which in the case of multiple populations of transiently heated grains can be a computationally costly operation \citep{1986A&A...160..295D, 2001ApJ...551..807D, 2015A&A...580A..87C}. Another example is the calculation of the level populations for atoms or molecules in the frame of Monte Carlo non-local thermodynamic equilibrium line radiative transfer \citep{2023A&A...678A.175M}. For both of these reasons, minimising the number of grid cells for a given resolution is thus beneficial for the simulation run time. 

Unstructured meshes also have some drawbacks. They are generally more complex to build and store than hierarchical grids and require more memory usage. In the context of Monte Carlo radiative transfer, the grid traversal time is particularly important. Grid traversal through a hierarchical grid is generally fast and it can be optimised using dedicated methods \citep{2013A&A...554A..10S}. For Voronoi grids the traversal is relatively slow since the grid cells have many faces, and each of these faces needs to be tested to determine the exit point for a given photon path \citep{2013A&A...560A..35C}. For tetrahedral grids, grid traversal is significantly faster than in Voronoi grids. In a tetrahedral grid, each cell has obviously only four faces, compared to 16 faces on average for a grid cell in a 3D Voronoi mesh \citep{1994A&A...283..361V}. Moreover, efficient algorithms based on Pl\"ucker coordinates and Pl\"ucker products have been designed for grid traversal through tetrahedral grids. For the same number of cells, however, the traversal is still not as efficient as in hierarchical octree grids.

\subsection{Post-processing hydrodynamical simulations}

The main reason why unstructured grids are interesting in the context of Monte Carlo radiative transfer is linked to post-processing hydrodynamical simulations. Full-scale hydrodynamical simulations are a powerful tool for studying the evolution of nearly all astronomical objects, and radiative transfer post-processing of such simulation outputs is needed to generate realistic mock observables that can be compared to real observations. The increased popularity of radiative transfer on Voronoi grids is primarily driven by the emergence of Lagrangian hydrodynamical solvers that operate on such unstructured Voronoi grids. The AREPO code \citep{2010MNRAS.401..791S, 2020ApJS..248...32W} is the most widely used one, but other solvers have been developed \citep{2012ApJ...755....7D, 2015ApJS..216...35Y, 2016A&C....16..109V, 2017MNRAS.471.3577C, 2018A&C....23...40V} or are being developed \citep{2023arXiv230513380S}. It is beneficial to use the same grid for the hydrodynamical calculations and the radiative transfer post-processing. Voronoi-based radiative transfer post-processing has been applied to generate mock observations for galaxies generated from moving-mesh cosmological hydrodynamical simulations, such as IllustrisTNG and AURIGA \citep[e.g.][]{2019MNRAS.483.4140R, 2020MNRAS.497.4773S, 2021ApJ...916...39C, 2022MNRAS.510.3321P, 2023MNRAS.519.4920G, 2024MNRAS.527.6506B}. 

The hydrodynamical equations can also be solved on tetrahedral grids. There has been a long history of tetrahedral grids in computational fluid dynamics simulations, in particular in engineering and material science applications \citep{Biswas1994, Biswas1996, Lohner1992, Ng2009, Xie2014, Xie2022}. The advantage of tetrahedral meshes is that they are very flexible and suited to fit within complex shapes with hard boundaries, such as buildings, turbines, or airplane wings. In astrophysical hydrodynamics we usually have no strong fixed boundaries for the computational domain. Nevertheless, hydrodynamical simulations on tetrahedral meshes have also been explored for astrophysical applications \citep{1999A&A...348..233B, 2002AGUFMSH52A0492L, kulikov2020hydrodynamic, kulikov2021molecular, kulikov2021computational}. If one wants to post-process such simulation outputs using radiative transfer on the same grid as used for the hydrodynamical simulations, the methodology discussed in this paper is required.

Tetrahedral grids might also prove useful for post-processing moving-mesh hydrodynamical simulations. However, in Monte Carlo radiative transfer it is commonly assumed that all physical quantities, such as density, are uniform within each cell \citep{2011BASI...39..101W, 2013ARA&A..51...63S}. In finite-volume hydrodynamics, of which moving-mesh hydrodynamics is a special case, the hydrodynamical simulations are solved using Gudonov's method, that is, by considering Riemann problems at all cell boundaries \citep{Toro1997, 2010MNRAS.401..791S}. Rather than the total masses in each Voronoi cell, the densities at the mesh-generating points are fundamental properties, and the assumption of a uniform density within each cell (the standard assumption in radiative transfer post-processing) is not consistent with the hydrodynamics method. 

In principle, we can avoid this inconsistency by performing the radiative transfer on the dual Delaunay tetrahedral grid rather than on the original Voronoi grid. Instead of assuming a uniform density in each of the tetrahedra cells, we can interpolate the density at each point based on the density values at the vertices of each tetrahedron, which are, by definition, the generating points of the moving-mesh simulation. An approach along this line has been implemented for Cartesian grids in the original version of SKIRT by \citet{2003MNRAS.343.1081B}. Rather than assigning a uniform density in each cuboidal grid cell, we used trilinear interpolation based on the values of the density at the eight vertices. As a result, the optical depth along a given path increases as a cubic function rather than as a linear function of the distance travelled along the path within each cell. This increased the run time for grid traversal by a factor of three. On the other hand, for a fixed number of cells the accuracy of the density of the interpolated grid is much larger, which compensates the longer run times. 

The main drawback of applying a similar approach to tetrahedral grids is its complexity. A planar triangulation of $n$ points results in ${\cal{O}}(n)$ triangles, but this does not apply to higher dimensions, where it is super-linear with an ${\cal{O}}(n^2)$ complexity.

\section{Conclusions}
\label{sec:Conclusions}

The proper discretisation of Monte Carlo radiative transfer simulations is crucial for obtaining accurate results. However, it is an intricate and complicated problem, as the `ideal' spatial grid should satisfy different, sometimes conflicting, criteria. Research into the advantages and disadvantages of different types of grids has been an important topic in the framework of the SKIRT Monte Carlo radiative transfer code \citep[e.g.,][]{2003MNRAS.343.1081B, 2013A&A...560A..35C, 2013A&A...554A..10S, 2014A&A...561A..77S}.

Tetrahedral grids have long been used in engineering applications, and more recently have been used for ray tracing in computer graphics. Coupled with its use in fluid and hydrodynamical simulations, these grid are promising tools in the context of Monte Carlo radiative transfer. 

We have successfully implemented a working tetrahedral grid structure within SKIRT. Our implementation uses the popular TetGen quality tetrahedral mesh
generator \citep{Si2015}, while for the grid traversal we implemented a very efficient method based on Pl\"ucker coordinates and Pl\"ucker products \citep{shoemake1998plucker, maria:hal-01486575}. 

The main conclusions of this work are the following:
\begin{itemize}
\item 
Tetrahedral grids are a very flexible grid structure that can be applied in a variety of simulations. The grid can be adaptively constructed using an underlying input medium, directly imported from an external source, or built from particle data. 
\item 
The correct implementation of the grid construction and the grid traversal algorithm were validated using 2D radiative transfer benchmark problems \citep{2004A&A...417..793P, 2009A&A...498..967P}. The SKIRT results based on tetrahedral grids and on regular 2D cylindrical grids reproduce the benchmark results very well. 
\item 
We compared the performance of tetrahedral, octree, and Voronoi grids using a simple 3D model. We find that, with a constant cell count, grid traversal through an octree grid is significantly more efficient than through a tetrahedral grid, which in turn is more efficient than a Voronoi grid. On the other hand, the tetrahedral grid has a lower grid quality compared to the other grid types. Combining these results, we find that, for a given runtime, the tetrahedral grid is consistently a lower-quality option than the other grid types. 
\item 
Despite the relatively poor performance of tetrahedral grids compared to Voronoi and octree grids, tetrahedral grids remain an interesting option for radiative transfer simulations, in particular for post-processing hydrodynamical simulations run on tetrahedral or unstructured grids. 
\end{itemize}

\begin{acknowledgements}
B.V. acknowledges support by the Fund for Scientific Research Flanders (FWO-Vlaanderen, project 11H2121N). 
\end{acknowledgements}

\bibliography{mybib.bib}

\end{document}